\documentclass[conference]{IEEEtran}
\IEEEoverridecommandlockouts
\usepackage{cite}
\usepackage{amsmath,amssymb,amsfonts}
\usepackage{algorithmic}
\usepackage{graphicx}
\usepackage{textcomp}
\usepackage{xcolor}
\usepackage{braket}
\usepackage{subcaption}
\usepackage{multirow}
\usepackage{comment}
\usepackage{hyperref}

\def\BibTeX{{\rm B\kern-.05em{\sc i\kern-.025em b}\kern-.08em
    T\kern-.1667em\lower.7ex\hbox{E}\kern-.125emX}}

\makeatletter
\newcommand{\linebreakand}{%
  \end{@IEEEauthorhalign}
  \hfill\mbox{}\par
  \mbox{}\hfill\begin{@IEEEauthorhalign}
}
\makeatother
\begin{document}

\title{A Study on Stabilizer R\'enyi Entropy Estimation using Machine Learning\\
}

\author{

    \IEEEauthorblockN{1\textsuperscript{st} Vincenzo Lipardi}
    \IEEEauthorblockA{\textit{Department of Advanced Computing Sciences} \\
    \textit{Maastricht University}\\
    Maastricht, The Netherlands \\
    vincenzo.lipardi@maastrichtuniversity.nl}
    \and
    \IEEEauthorblockN{2\textsuperscript{nd} Domenica Dibenedetto}
    \IEEEauthorblockA{\textit{Department of Advanced Computing Sciences} \\
    \textit{Maastricht University}\\
    Maastricht, The Netherlands \\
    domenica.dibenedetto@maastrichtuniversity.nl
    }
    \linebreakand
    \IEEEauthorblockN{3\textsuperscript{rd} Georgios Stamoulis}
    \IEEEauthorblockA{\textit{Department of Advanced Computing Sciences} \\
    \textit{Maastricht University}\\
    Maastricht, The Netherlands \\
    georgios.stamoulis@maastrichtuniversity.nl
    }
    \and
    \IEEEauthorblockN{4\textsuperscript{th} Mark H.M. Winands}
    \IEEEauthorblockA{\textit{Department of Advanced Computing Sciences} \\
    \textit{Maastricht University}\\
    Maastricht, The Netherlands \\
    m.winands@maastrichtuniversity.nl
    }
    
    }

\maketitle

\begin{abstract}
Nonstabilizerness is a fundamental resource for quantum advantage, as it quantifies the extent to which a quantum state diverges from those states that can be efficiently simulated on a classical computer, the stabilizer states. The stabilizer R\'enyi entropy (SRE) is one of the most investigated measures of nonstabilizerness because of its computational properties and suitability for experimental measurements on quantum processors. Because computing the SRE for arbitrary quantum states is a computationally hard problem, we propose a supervised machine-learning approach to estimate it.
In this work, we frame SRE estimation as a regression task and train a Random Forest Regressor and a Support Vector Regressor (SVR) on a comprehensive dataset, including both unstructured random quantum circuits and structured circuits derived from the physics-motivated one-dimensional transverse Ising model (TIM). We compare the machine-learning models using two different quantum circuit representations: one based on classical shadows and the other on circuit-level features. Furthermore, we assess the generalization capabilities of the models on out-of-distribution instances. Experimental results show that an SVR trained on circuit-level features achieves the best overall performance. On the random circuits dataset, our approach converges to accurate SRE estimations, but struggles to generalize out of distribution. In contrast, it generalizes well on the structured TIM dataset, even to deeper and larger circuits. In line with previous work, our experiments suggest that machine learning offers a viable path for efficient nonstabilizerness estimation.
\end{abstract}


\section{Introduction}
Nonstabilizerness, also known as \textit{magic}, is a property of quantum states that measures their hardness to be classically simulated~\cite{nielsen2010quantum}, and thus it constitutes a fundamental resource for quantum advantage~\cite{gottesman1998theory,campbell2017roads, howard2017application}. According to the Gottesman-Knill Theorem~\cite{gottesman1998theory}, quantum computations involving exclusively Clifford operations can be efficiently simulated on a classical computer~\cite{nielsen2010quantum}. Quantum states that can be prepared using exclusively Clifford operations are called stabilizer states.
Hence, nonstabilizerness quantifies to what extent a quantum state diverges from the set of stabilizer states.
Notably, stabilizer states can exhibit full entanglement, which means that entanglement alone is not a sufficient indicator of quantum advantage. For example, Bell states are maximally entangled but are also stabilizer states, as they can be prepared using Clifford gates only (X, H, and CNOT). Furthermore, Clifford gates alone are not sufficient to form a universal gate set. Therefore, studying the role of non-Clifford operations in quantum algorithms is central when studying quantum advantage.

Several measures of nonstabilizerness have been recently proposed~\cite{veitch2012negative}, such as stabilizer nullity~\cite{beverland2020lower}, robustness of magic~\cite{heinrich2019robustness, howard2017application}, Bell magic~\cite{haug2023scalable}, and stabilizer R\'enyi entropies (SREs)~\cite{leone2022stabilizer}.
This work focuses on the broadly used SRE, because of its favorable computational properties~\cite{ahmadi2024quantifying, leone2022stabilizer}, and its practical suitability for experimental measurement on quantum hardware~\cite{oliviero2022measuring}. However, the computational cost to compute the SRE in general grows exponentially with respect to the number of qubits~\cite{leone2022stabilizer}. 
Tensor Networks~\cite{orus2014practical} are one of the most investigated and promising approach to address this challenge~\cite{haug2023quantifying, lami2023nonstabilizerness, lami2024unveiling, tarabunga2023many}. However, they have restricted application to quantum states with weak entanglement and low dimensionality. A more general approach is based on neural quantum states \cite{sinibaldi2025non}. Recent research has also explored Machine Learning (ML) models to classify stabilizer states~\cite{mello2025retrieving}. In particular, Convolutional Neural Networks~\cite{mello2025retrieving} have been proposed to distinguish between stabilizer and nonstabilizer states. Although this approach achieves interesting results, including prediction that are invariant under the Clifford group, it is limited to the classification formulation of the problem. 

 
In this paper, we propose supervised ML models to tackle the SRE estimation formulated as a regression task. The primary goal is to reduce the runtime complexity by trading off online computational cost for an approximated SRE estimation.
The main contributions of the paper are as follows.
\begin{enumerate}
    \item We generate and release a comprehensive dataset of quantum circuits for the SRE estimation task. It comprises two classes of circuits: unstructured random quantum circuits and structured circuits based on the one-dimensional Transverse Ising Model. 
    \item We train two supervised ML models for SRE estimation, including the Random Forest Regressor and the Support Vector Regressor. Performance is evaluated both in-distribution, on circuits structurally similar to the training data, and out-of-distribution, including circuits with increased gate count and qubit number~\cite{liu2021towards}. 
    \item We investigate two classical representations of quantum circuits as model inputs. The first is a circuit-level representation based on gate counts~\cite{liao2024machine}. The second is derived from classical shadows~\cite{huang2020predicting, huang2023learning, huang2022provably}, constructed by measuring the expectation values over 1- and 2-local observables. Additionally, we analyze a combined feature set that integrates both representations.
\end{enumerate}

This paper is organized as follow. Section \ref{sre} introduces the stabilizer R\'enyi entropy. Section \ref{data_generation} presents the dataset generation protocol. Section \ref{methods} describes the experimental setup and ML models. Section \ref{results} reports the results collected in the different experimental settings. Finally, Section \ref{conclusion} presents the main outcomes of this study and future research. All data and code to reproduce our results are publicly available at \url{https://github.com/VincenzoLipardi/SRE-Estimation}.

\section{Stabilizer R\'enyi Entropy}\label{sre}
The \textit{Stabilizer R\'enyi Entropy} (SRE)~\cite{leone2022stabilizer} of order $\alpha$ for a pure $n$-qubit quantum state $\rho$ is defined as:
\begin{equation}
    S_{\alpha} (\rho)= \frac{1}{1-\alpha} \log \sum_{P\in \mathcal{P}_n}\Xi^{\alpha}_P(\rho) -\log(2^n) \label{sre_formula}
\end{equation}

\noindent where $\mathcal{P}_N$ denotes the set of $n$-qubit Pauli strings and $\Xi_P(\rho)= \frac{1}{2^n}Tr(\rho P)^2$. A \textit{Pauli string} is a tensor product of Pauli matrices, each acting on a different qubit in a multi-qubit system. When $\rho$ is a stabilizer state, $S_{\alpha}(\rho) = 0$ for all $\alpha \geq 2$. According to previous works we fix $\alpha=2$~\cite{leone2024stabilizer, ahmadi2024quantifying, sinibaldi2025non}. 

In general, calculating the SRE of a quantum state using the Equation~\ref{sre_formula} is hard, as it requires estimating $4^n$ expectation values, corresponding to all possible combinations of Pauli strings. 


\section{Data Generation} \label{data_generation}
The dataset comprises two classes of quantum circuits, both with the number of qubits \( n \) ranging from 2 to 6. The first class includes \(50{,}000\) random quantum circuits, \(10{,}000\) for each value of \( n \), and is referred to as the RQC dataset. The second class includes \(5{,}000\) structured circuits based on the Trotterized dynamics of the one-dimensional Ising model, \(1{,}000\) for each value of \( n \), and is referred to as the TIM dataset. The Ising model has been chosen for its fundamental importance across a broad range of fields, including condensed matter physics, statistical mechanics, and more. All circuits are labeled with their SRE value, calculated as described in Section~\ref{sre}.
These datasets are designed to be representatives of typical circuits used in applications for NISQ devices~\cite{preskill2018quantum}, covering a wide range of SRE values. This variety is crucial for studying the strengths and limitations of our approach.

\begin{figure*}[!ht]
    \centering
    \begin{subfigure}[t]{0.49\textwidth}
        \centering
        \includegraphics[width=\linewidth]{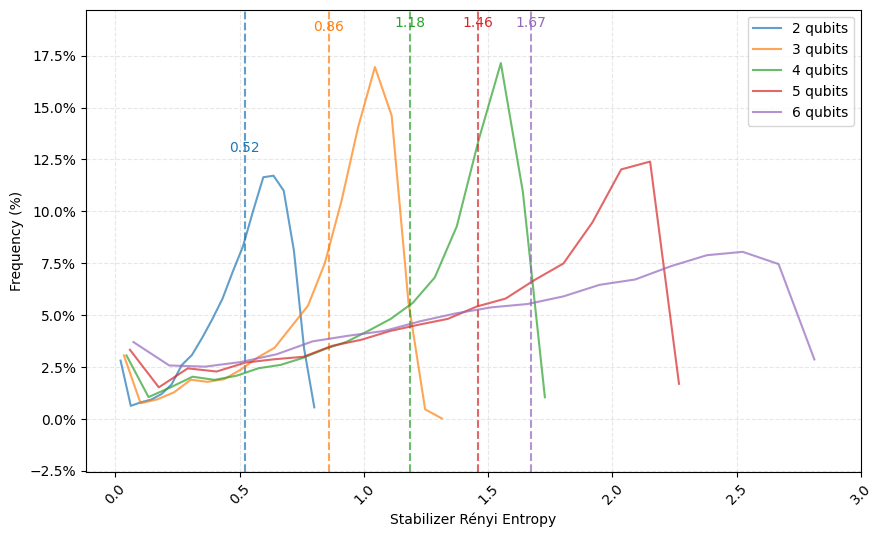}
        \caption{RQC Dataset - SRE distribution per qubit}
        \label{fig:random}
    \end{subfigure}
    \hfill
    \begin{subfigure}[t]{0.49\textwidth}
        \centering
        \includegraphics[width=\linewidth]{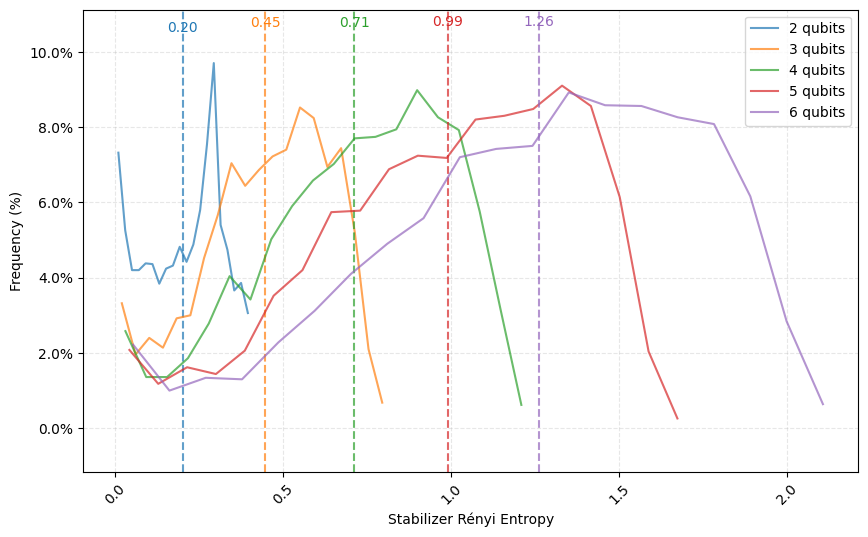}
        \caption{TIM Dataset - SRE distribution per qubit}
        \label{fig:ising}
    \end{subfigure}
    
    \vspace{0.1cm}
    
    \begin{subfigure}[t]{0.49\textwidth}
        \centering
        \includegraphics[width=\linewidth]{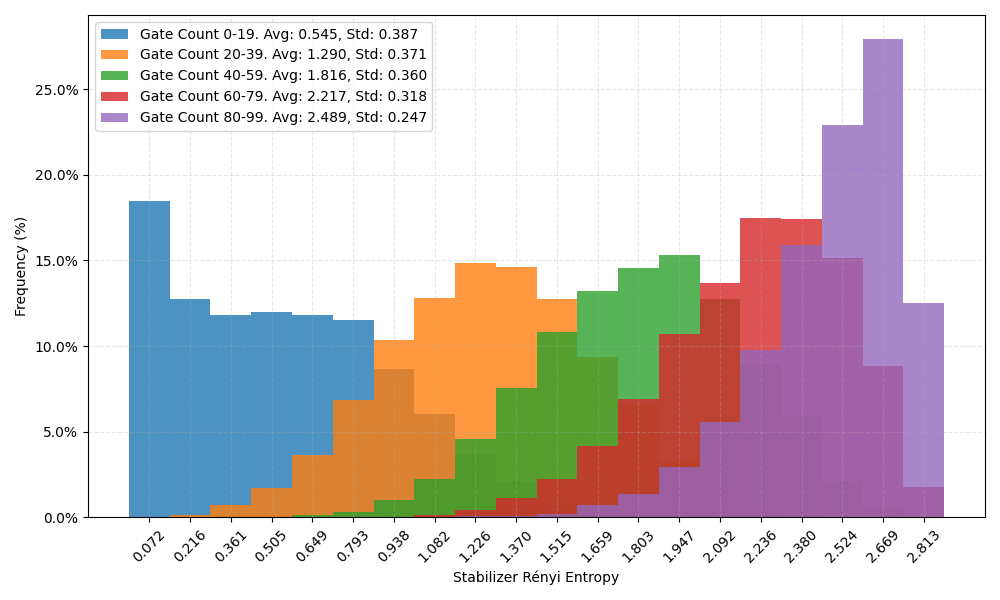}
        \caption{RQC Dataset- SRE distribution for 6-qubit circuits}
        \label{fig:random_6}
    \end{subfigure}
    \hfill
    \begin{subfigure}[t]{0.49\textwidth}
        \centering
        \includegraphics[width=\linewidth]{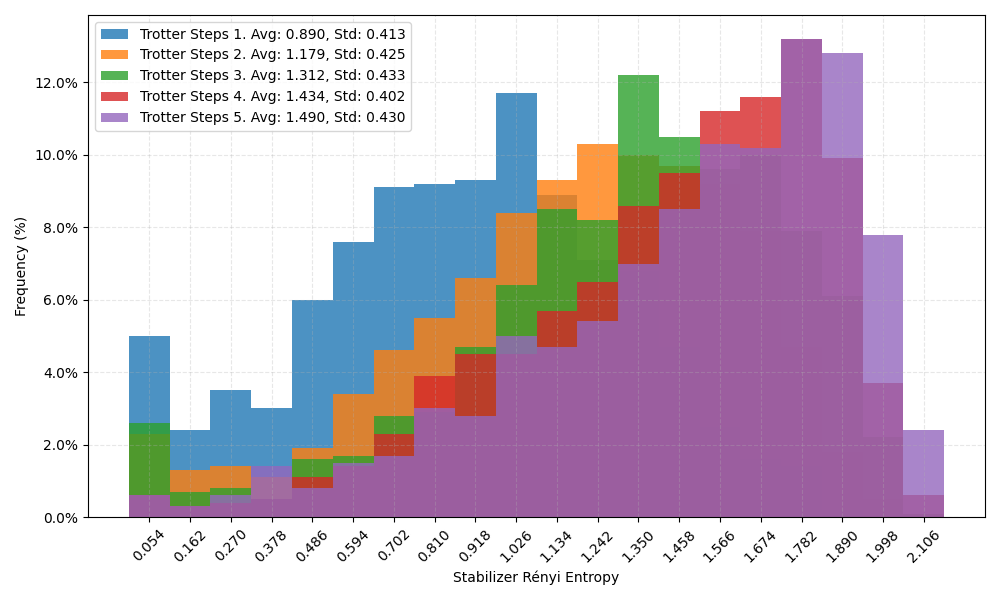}
        \caption{TIM Dataset - SRE distribution for 6-qubit circuits}
        \label{fig:tim_6}
    \end{subfigure}
    
    \caption{
        The frequency of the SRE values of the quantum circuits, grouped by the number of qubits, in the RQC and TIM dataset, in Figures~\ref{fig:random} and ~\ref{fig:ising}, respectively. 
        Figures~\ref{fig:random_6} and~\ref{fig:tim_6} show the frequency of the SRE values of the 6-qubit circuits, grouped by the number of gates and trotter steps.}
    \label{fig:data}
\end{figure*}

Figure \ref{fig:data} illustrates the SRE frequency for RQC and TIM datasets. Subplots \ref{fig:random} and \ref{fig:ising} show SRE frequencies across qubit counts, while \ref{fig:random_6} and \ref{fig:tim_6} show SRE distributions for 6-qubit circuits, grouped by gate counts and Trotter steps, respectively. 

In the RQC dataset, once we fix the number of qubits $n$, the circuits are generated by randomly sampling the number of gates uniformly from the range \(G=[0, 100]\). Starting with an empty quantum circuit, gates are sequentially added. Each gate is randomly picked from a universal gate set that includes the \texttt{CNOT} gate and the three single-qubit rotation gates: \texttt{RX}, \texttt{RY}, \texttt{RZ}.

In the TIM dataset, each circuit encodes a discrete-time approximation of the Transverse Ising Hamiltonian on a chain:
\[
H = -J \sum_{i=1}^{n-1} Z_i Z_{i+1} - h \sum_{i=1}^{n} X_i,
\]
where \( Z_i \) and \( X_i \) denote Pauli operators acting on qubit \( i \), \( J \) is the interaction strength, and \( h \) is the transverse field strength.
We simulate the time evolution operator \( U(t) = e^{-i H t} \) using a first-order Trotter-Suzuki decomposition~\cite{nielsen2010quantum}. Each Trotter step is implemented as a sequence of gates in the quantum circuit as follows:
\begin{itemize}
  \item The two-qubit interaction \( e^{-i \theta Z_i Z_{i+1}} \), with \( \theta = J \Delta t \), is implemented using a \texttt{CNOT}–\texttt{RZ}–\texttt{CNOT} sequence:
  \[
  \texttt{CNOT}_{i,i+1} \rightarrow \texttt{RZ}(2\theta)_{i+1} \rightarrow \texttt{CNOT}_{i,i+1}
  \]
  \item The transverse-field term \( e^{-i \phi X_i} \), with \( \phi = h \Delta t \), is implemented via a single-qubit rotation
  $
  \texttt{RX}(2\phi)_{i}$.
  
\end{itemize}

These circuits span various values for the angle parameters $\theta$, $\phi$, with the number of trotter steps ranging in \(T=[1, 5]\).

Subsection~\ref{circuit-level} and Subsection~\ref{classical_shadows} introduce the two different classical representations used for the quantum circuits in the dataset.

\subsection{Circuit-level Feature Encoding} \label{circuit-level}
The circuit-level input features are extracted directly from the quantum circuits as counts of each gate type. Following~\cite{liao2024machine}, parameterized gates are counted using binned rotation angles, where the interval $[0, 2\pi]$ is divided into $50$ bins. This results in a total of $152$ input features. These features are then used as input for the RFR and  SVR models. 

\subsection{Shadow-based Feature Encoding} \label{classical_shadows}
The protocol of classical shadows can be used to predict any number of properties of a quantum state $\rho$ using a logarithmic number of measurements with respect to the number of properties~\cite{huang2020predicting}. These properties are the expectation values of observables $O_i$ on the quantum state $\rho$:
\begin{equation}
    o_i = Tr(O_i \rho), \hspace{0.5cm} 1\leq i\leq M.\label{exp_val}
\end{equation}

For each circuit in the two datasets, we compute the classical shadow with respect to a set of observables consisting of all Pauli strings that act non-trivially on at most two qubits. Consequently, the total number of classical shadows computed, denoted by \( F(n) \), depends on the number of qubits \( n \):

\begin{equation}
    F(n) = 3n + 9 \binom{n}{2}. \label{shadows_feature_number}
\end{equation}

In Equation~\ref{shadows_feature_number}, the first term  accounts for all single-qubit Pauli operators (i.e., \( X, Y, Z \) on each qubit), while the second term  accounts for all two-qubit Pauli strings that act non-trivially on exactly two distinct qubits (i.e., \( 3 \times 3 = 9 \) combinations of Pauli operators for each qubit pair).
The use of classical shadows in machine learning is particularly promising~\cite{huang2023learning}, as highlighted in the field of condensed matter physics~\cite{huang2022provably}.\\
The classical shadow protocol is divided into two main parts. The first is the data collection based on randomized measurements, while the second is a classical post-processing of the data meant to reconstruct the quantum state. Specifically, the classical shadow of a given state $\rho$ is obtained by performing the following steps iteratively. First, the data collection, which samples a unitary $U_i$ from a predefined ensemble of efficient-to-simulate unitaries and apply it to the quantum state, $ \rho\rightarrow U_i^{\dagger} \rho U_i$; then measure all qubits in the computational basis and obtain a bit string $\ket{b_i}$. Second, the classical postprocessing, which inverts the measurement channel $\mathcal{M}(\rho) = \mathbb{E} \left[ U_i^{\dagger} \ket{b_i}\bra{b_i}U_i \right]$.

The second part can be performed because the unitaries $U_i$ are known and efficient to simulate. After $N$ iterations, we obtain $N$ single classical snapshots of $\rho$:
\begin{equation}
    \hat{\rho}_i = \mathbb{E} \left[ \mathcal{M}^{-1} (U_i^{\dagger} \ket{b_i}\bra{b_i}U_i)\right]  \label{shadow}
\end{equation}
The collection of all $\hat{\rho}_i $ is referred to as \textit{classical shadow} of the state $\rho$ and $N$ as its \textit{size}.
A classical shadow of size $N$ can predict $M$ properties in the form of Eq. \ref{exp_val} within an additional error $\epsilon$ when the following relation is fulfilled
\begin{equation}
    N\geq \mathcal{O}\left( \frac{\log (M) \max_i \lVert O_i\rVert_{shadow}}{\epsilon^2}\right).
\end{equation}
Hence, all the properties that are linear in $\rho$ as in the Equation \ref{exp_val} can be predicted efficiently in the sample complexity if the observables have finite norm $\lVert \cdot \rVert _{shadow} $\cite{huang2020predicting}. 
Note that the norm $\lVert \cdot \rVert _{shadow} $ depends on the unitary ensemble chosen. For computational reasons, we chose the set of unitaries obtained as tensor products of random single-qubit Clifford circuits. In this case, the norm scales exponentially in the locality of the observable $\lVert O_i \rVert _{shadow}\leq 4^k \lVert O_i\rVert _{\infty}$, where $k$ is the maximum number of qubits on which $O_i$ acts non-trivially. 


\section{Machine-Learning Models} \label{methods}
In this section, we present the machine learning models, in Section~\ref{random_forest} and Section~\ref{support_vector_machine}, used to estimate the SRE from the quantum circuit representations described above. We evaluate their performance on both interpolation and extrapolation tasks, with a focus on generalization across circuit size and structure, as discussed in Section~\ref{inter_extra}.

\subsection{Random Forest Regressor} \label{random_forest}
Random Forest Regressor (RFR) is a non-linear ensemble learning method based on aggregating the outputs of multiple decision trees~\cite{breiman2001random}. We choose RFR for its robustness, explainability, and strong performance on tabular datasets with complex, non-linear feature interactions. We perform a grid search as a hyperparameter selection strategy with cross-validation to tune the number of estimators (trees), the maximum tree depth, and the splitting criterion. 

\subsection{Support Vector Regressor}\label{support_vector_machine}
Support Vector Regression (SVR) is a kernel-based method that constructs a regression function by fitting a subset of training data, known as support vectors, within a specified margin~\cite{drucker1996support}.
We choose SVR for its robustness to outliers and strong performance on small- to medium-sized datasets, owing to its capacity to model non-linear relationships through kernel functions. 
We perform a grid search as a hyperparameter selection strategy with cross-validation to tune the regularization parameter, the margin, and the kernel type.

\subsection{Interpolation and Extrapolation} \label{inter_extra}
To assess the models, the experiments are divided into two main parts. The first focuses on \textit{interpolation}, where we evaluate the capability of the models to predict SRE on in-distribution circuits, which are unseen during the training phase by the model but share similar structural characteristics, such as the number of qubits and of total gates count. The second part focuses on the \textit{extrapolation}, where we evaluate the capability of the models to generalize the predictions to out-of-distribution circuits, which lay outside the support of the training distribution. Specifically, we study the extrapolation over circuits with a higher number of qubits and of total gates than those seen during training. The extrapolation part is a challenging out-of-distribution generalization task~\cite{liu2021towards}.

\section{Results} \label{results}
To assess the performance of our models we perform a runtime analysis, in Section \ref{runtime_analysis}, and assess the quality of the SRE estimations based on the model's interpolation and extrapolation capabilities, in Section \ref{interpolation} and Section \ref{extrapolation} respectively. 
In all the experiments, we evaluate the SRE estimation based on the Mean Squared Error (MSE). Moreover, each dataset is randomly partitioned into training and test sets (80\%–20\%).

\subsection{Runtime Analysis}\label{runtime_analysis}
\begin{figure}[!t]
    \centering
    \includegraphics[width=1\linewidth]{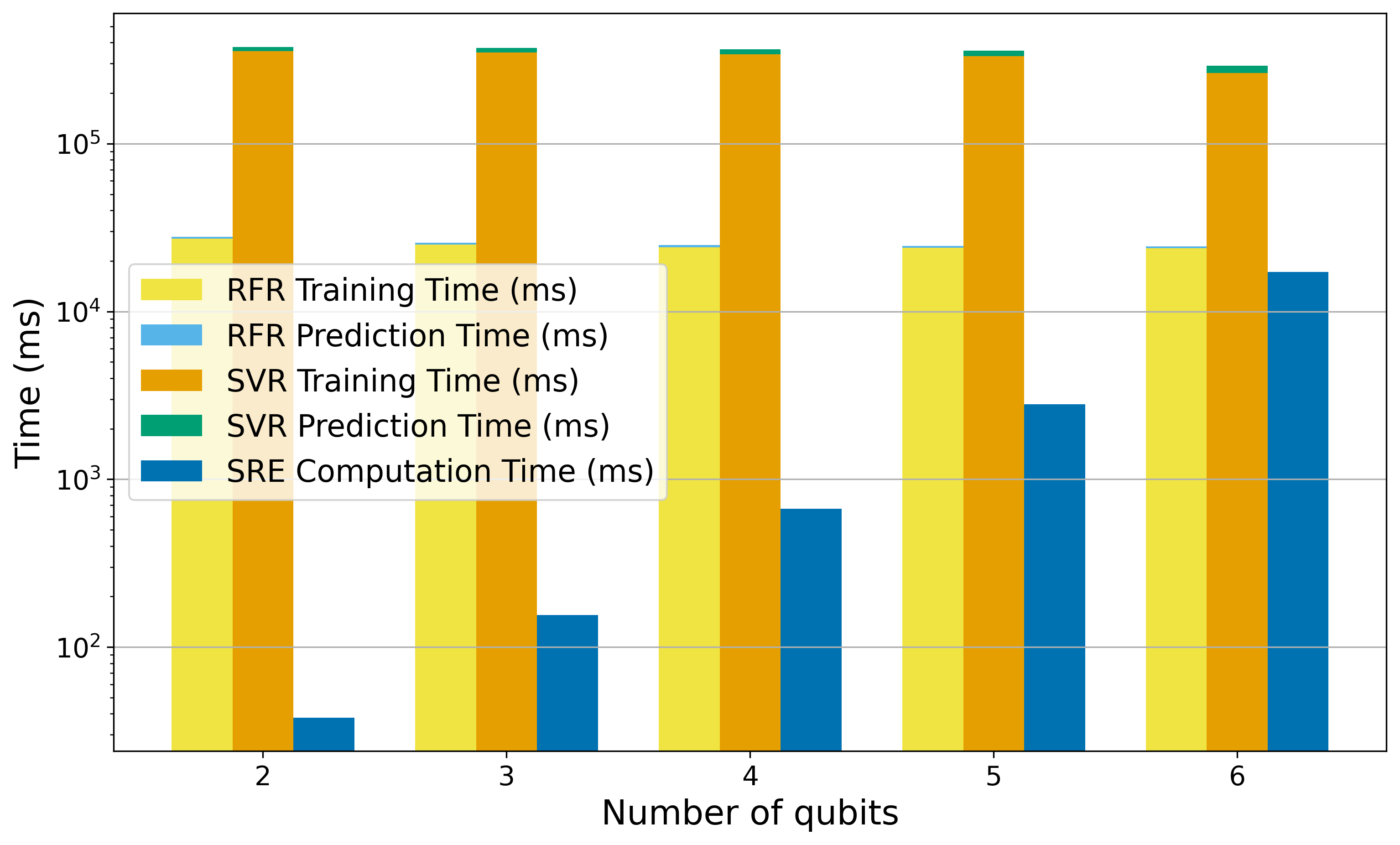}
    \caption{Comparison of training and prediction times of the ML models with the runtime of an exact SRE computation. }
    \label{fig:time_analysis}
\end{figure}

Training and prediction times for each machine learning model are compared with the time required to compute the SRE for a single circuit using Equation~\ref{sre_formula}. Figure~\ref{fig:time_analysis} presents this comparison, with the number of qubits on the $x$-axis and runtime (in milliseconds) on the $y$-axis, shown on a logarithmic scale.
To have a fair comparison, the time values in blue are calculated averaging over $50$ random circuits sampled from the RQC dataset with gate range $G\in[40,59]$. In terms of computational time, the ML models exhibit a clear advantage. The runtime required to calculate the SRE on quantum circuits grows exponentially with the number of qubits, whereas the training time of the ML models remains constant. Moreover, the prediction time, which is the relevant metric after the one-time training, is negligible compared to the time needed to compute the SRE, even for a simple two-qubit quantum circuit.

\begin{figure*}[!ht]
\vspace{-0.5em}
    \centering
    \begin{subfigure}[t]{0.48\textwidth}
        \centering
        \includegraphics[width=\linewidth]{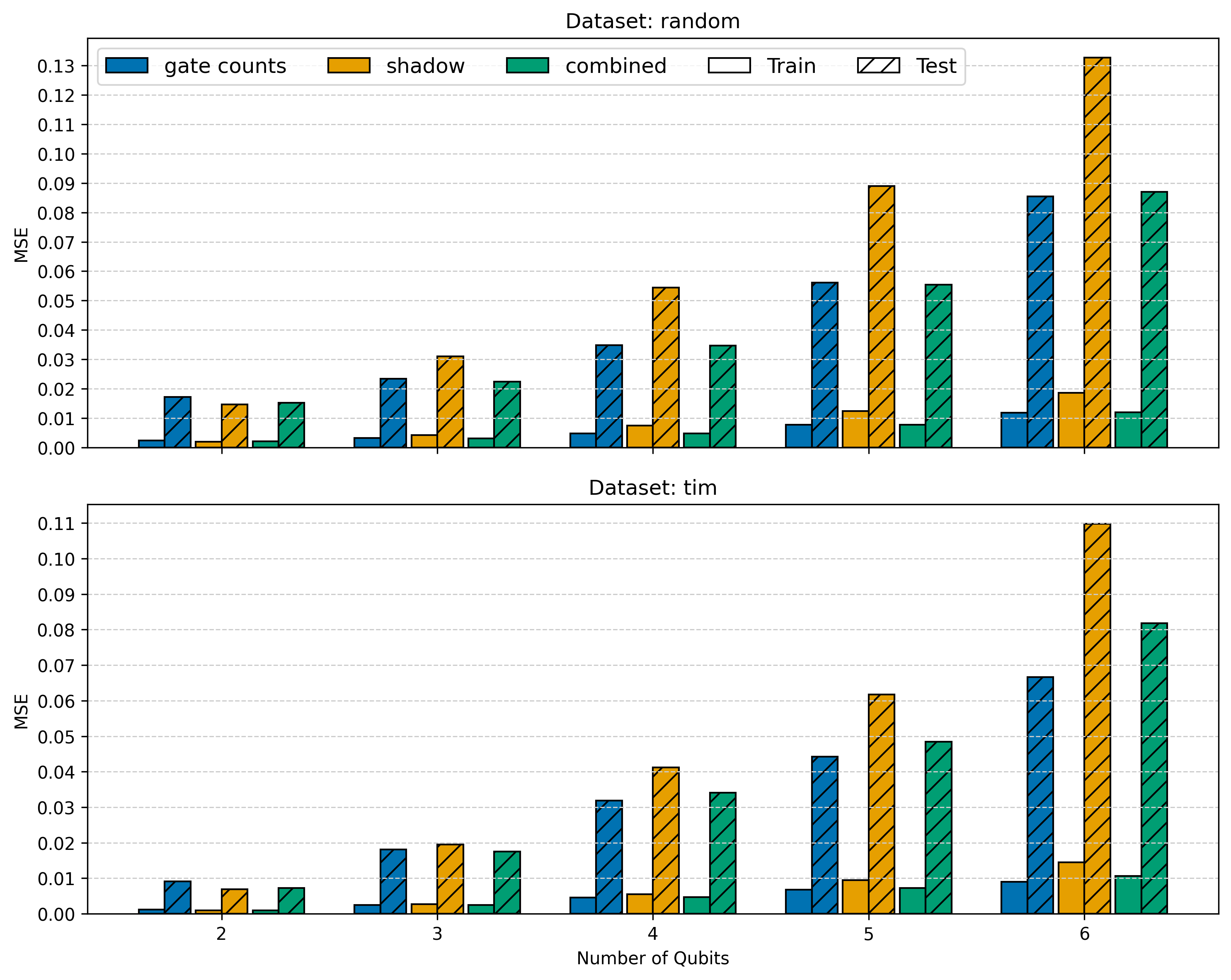}
        \caption{Random Forest Regressor}
        \label{fig:rfr_indep}
    \end{subfigure}
    \hfill
    \begin{subfigure}[t]{0.48\textwidth}
        \centering
        \includegraphics[width=\linewidth]{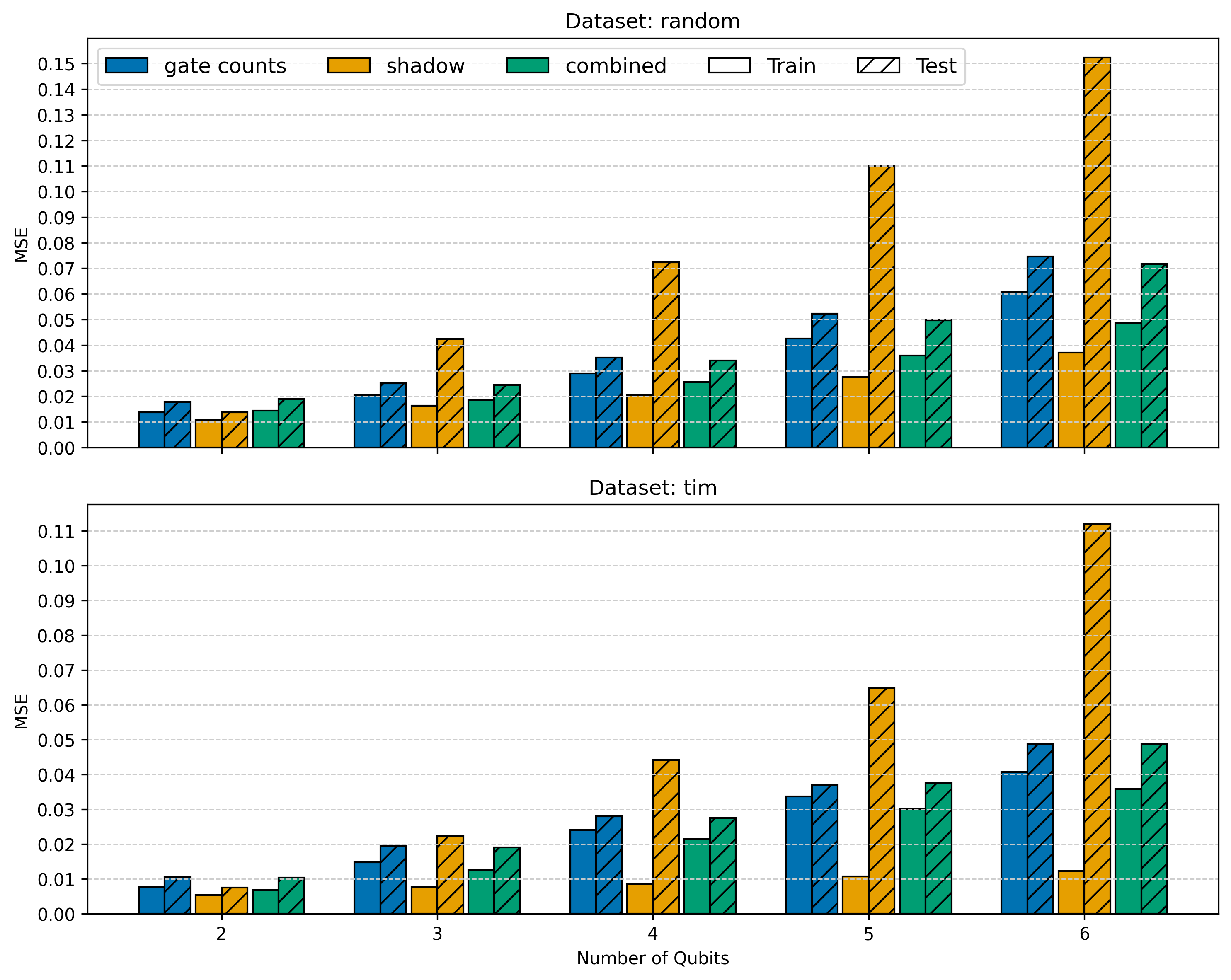}
        \caption{Support Vector Regressor}
        \label{fig:svr_indep}
    \end{subfigure}
    \caption{Mean Squared Error (MSE) performance of the ML models in the interpolation scenario.
}
    \label{fig:results_independent}
    
\end{figure*}
\begin{figure*}[!ht]
    \centering
    \begin{subfigure}[t]{0.48\textwidth}
        \centering
        \includegraphics[width=\linewidth]{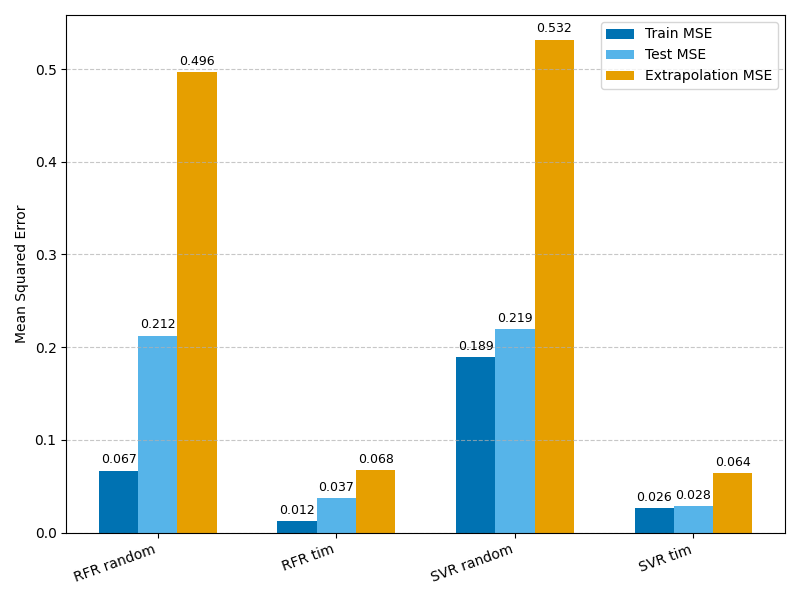}
        \caption{Extrapolation in circuit depth}
        \label{fig:extrapolation_depth}
    \end{subfigure}
    \hfill
    \begin{subfigure}[t]{0.48\textwidth}
        \centering
        \includegraphics[width=\linewidth]{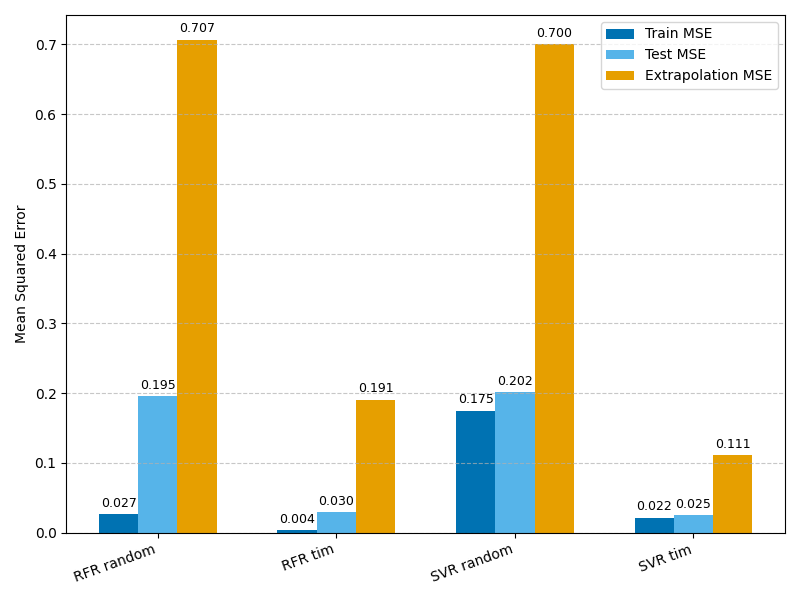}
        \caption{Extrapolation in qubit count}
        \label{fig:extrapolation_qubits}
    \end{subfigure}
    \caption{Mean Squared Error (MSE) performance of the ML models in the extrapolation scenario.}
    \label{fig:extrapolation}
    \vspace{-0.5em}
\end{figure*}

\subsection{Interpolation}\label{interpolation}
In these experiments, the models are trained independently on quantum circuits with fixed number of qubits. 
Figure~\ref{fig:results_independent} illustrates the interpolation performance of RFR ~\ref{fig:rfr_indep} and SVR ~\ref{fig:svr_indep} across the three different feature sets on both the RQC and TIM datasets, in the top and bottom row respectively. RFR and SVR fit the data with low train MSE (solid bars) and test MSE (hatched bars), and especially on the TIM dataset. Compared to SVR, RFR obtains MSE values with higher discrepancy between the training and test set, suggesting that SVR is able to capture better patterns in the data and that RFR may be affected by overfitting. Notably, SVR coupled with circuit-level features presents small difference between train and test MSE, confirming good generalization performance. 

Across both datasets and for both models, the circuit-level representation (blue bars) provides a reasonable baseline. On the RQC dataset, the test set MSE remains below $0.09$ for RFR and below $0.08$ for SVR. While on the TIM dataset, it stays below $0.07$ for RFR and below $0.05$ for SVR.
Classical shadows (orange bars) consistently achieve better performance on the training set and exhibit more favorable MSE scaling compared to the other feature sets. However, models trained on classical shadows present lower generalization performance.
The combined feature vector (green bars) paired with SVR consistently has the lowest test MSE, indicating that classical shadows provide complementary information to the simpler circuit-level statistics. 
We note that, on average, the SRE value of quantum circuits increases with the number of qubits, as shown in Figure \ref{fig:data}. Therefore, an increase in MSE values with the number of qubits is expected, even under constant relative errors.

\subsection{Extrapolation}\label{extrapolation}
To evaluate the generalization capabilities of our models, we perform two types of extrapolation experiments: qubit and gate count. In the qubit extrapolation experiments, the models are trained on the subset of quantum circuits with qubit numbers from $2$ to $5$, and then evaluated on $6$-qubit circuits. In the gate count extrapolation, the models are trained on the subset of circuits with gate counts restricted to the range $0-79$ for the RQC dataset and evaluated on circuits with gate counts ranging in $80-99$. This approach is similarly applied to the TIM dataset, where the model is trained on circuits with trotter steps 1-4 and tested on step 5.\\
Figure~\ref{fig:extrapolation} illustrates the extrapolation performance of RFR and SVR for estimating the SRE in both scenarios.
In the depth‐extrapolation setting (\subref{fig:extrapolation_depth}) both RFR and SVR properly fit the data, achieving test set MSE values below $0.2$ and $0.04$ on the RQC and TIM datasets, respectively. However, the extrapolation set MSE increases by factors $2.5$ and $2$ on the two datasets. As expected, the models have significantly better performance on the structured data underlying the symmetries of the 1D transverse field Ising model. 
In the qubit‐extrapolation setting (\subref{fig:extrapolation_qubits}) there is a similar pattern. The test set MSE of $0.2$ and $0.03$, increases by factor $3.5$ and $4-6$ on the RQC and TIM datasets, respectively. Overall, the SVR model demonstrates superior performance, exhibiting a lower tendency to overfit and better generalization capabilities.

\section{Conclusion and Future Research} \label{conclusion}
This paper investigates ML models to estimate SRE for quantum states~\cite{leone2022stabilizer}. In line with previous research~\cite{mello2025retrieving}, we identify ML as a promising approach to address SRE estimation. Our results show that ML models offer an alternative to conventional methods by trading off online computational cost for an approximated SRE estimation, avoiding the exponential runtime growth with qubit number.

We observe that both Random Forest Regressor (RFR) and Support Vector Regressor (SVR) can successfully fit the training data, particularly when enhanced with a classical shadow representation. However, these models generally exhibit limited generalization performance, especially on out-of-distribution instances in the RQC dataset. Among the two, SVR performs best on both in-distribution and out-of-distribution instances, achieving notable results on the TIM dataset with mean squared errors of $0.06$ and $0.1$, respectively, when generalizing to unseen circuit depths and numbers of qubits. Interestingly, models trained on classical shadows show lower training error, while those using circuit-level features generalize better.
We note that these models may prove useful in real-time applications, where strict time constraints require fast and thus approximated SRE estimations. For example, in quantum architecture search~\cite{lipardi2025quantum} it is fundamental to keep the quantum circuit simulation time low, but SRE estimation can be considered to guide the search towards hard-to-simulate quantum circuits.

There are two main directions for future research. First, the design of different efficient but more informative quantum circuit representations may improve the performance of our models. In this direction, more complex models like graph neural networks (GNNs), which naturally encode the circuit topology using the directed acyclic graph representation~\cite{liao2024machine, aktar2024graph}, could capture the most relevant features to generalize more effectively. GNNs are also well-suited to embedding hardware-specific information, making them promising candidates for extending SRE estimation to real quantum hardware. Second, incorporating these models into the quantum architecture search framework to enable the development of SRE-informed techniques. This may be crucial to achieve a quantum advantage, as the resulting quantum circuits will be designed by considering both the quality of the solution provided to the target problem and the hardness of simulating the circuit on a classical computer.

\bibliographystyle{IEEEtran}

\bibliography{main}

\end{document}